\documentstyle[prb,aps,preprint]{revtex}

\begin{document}
\draft
\title{A scalable solid-state quantum computer based on quantum dot pillar 
structures}
\author{G. D. Sanders, K. W. Kim, and W. C. Holton}
\address{Department of Electrical and Computer Engineering,
North Carolina State University\\
Raleigh, North Carolina 27695-7911}

\maketitle
\begin{abstract}

We investigate an optically driven quantum computer based on electric dipole 
transitions within coupled single-electron quantum dots.
Our quantum register consists of a 
freestanding n-type pillar containing a series of pair wise coupled 
asymmetric quantum dots, each with a slightly different energy structure, 
and with grounding leads at the top and bottom of the pillar. Asymmetric 
quantum wells confine electrons along the pillar axis and a negatively 
biased gate wrapped around the center of the pillar allows for electrostatic 
confinement in the radial direction. We self-consistently solve 
coupled Schr\"{o}dinger and Poisson equations and develop 
a design for a three-qubit quantum register. Our results indicate that a 
single gate electrode can be used to localize a single electron in each 
of the quantum dots. Adjacent dots are strongly coupled by electric 
dipole-dipole interactions arising from the dot asymmetry, thus enabling 
rapid computation rates. The dots are tailored to minimize dephasing due to 
spontaneous emission and phonon scattering and to maximize the number of 
computation cycles. The design is scalable to a large number of qubits.

\end{abstract}



\section{Introduction}
 
The possibility that a computer with exceptional properties could be built 
employing the laws of quantum physics has stimulated considerable interest 
in searching for useful algorithms and realizable physical implementations.
Two useful algorithms, exhaustive search \cite{bib:Grover97} and 
factorization, \cite{bib:Shor94} have been discovered; others, including 
the suggestion that quantum computers will prove useful to model quantum 
systems, are being sought. Meanwhile, various physical implementations are 
being explored, including trapped ions, \cite{bib:Cirac95} cavity quantum 
electrodynamics, \cite{bib:Pellizzari95} ensemble nuclear magnetic resonance,
\cite{bib:NMR}  small Josephson junctions, \cite{bib:Shnirman97} 
optical devices incorporating beam splitters and phase shifters
\cite{bib:Cerf98} and a number of solid state systems based on quantum dots.
\cite{bib:BarencoDeutsch95,bib:Wang,bib:Kane,bib:Tanamot099,bib:Loss98}
Although the advantages of quantum computing are enormous for particular 
key applications, the requirements for their implementation are extremely 
stringent, perhaps especially rigorous for solid-state systems. 
Nevertheless solid-state quantum computers are very appealing relative 
to other possible implementation schemes because of the well-known ability 
to customize the design through the use of artificially structured materials 
and the probable scalability of the resulting design. 
For example, integrated circuit manufacturing technology would be 
immediately applicable to quantum computers of the proper implementation; 
and such designs would not only be scalable to smaller dimensions along 
the "semiconductor learning curve" but also large ensembles of "identical" 
quantum computers could be manufactured, that could be individually 
fine-tuned electrically. To date, no solid-state implementation of quantum 
computing has been demonstrated.

In this paper, we investigate a solid-state quantum computer implementation 
that is amenable to manufacturing with integrated circuit technology. 
We develop a three-dimensional (3D) device model and self-consistently solve 
coupled Schr\"{o}dinger and Poisson equations to
generate a quantum computer design for a three-qubit quantum register 
that is based on pair wise coupled asymmetric III-V quantum dots. 
The design is optimized for a long coherence time and a rapid computation 
rate. Our results indicate that this structure may provide a realistic 
scalable candidate for quantum computing in solid-state systems.

\section{Proposed Structure}

The proposed quantum dot quantum computer (see Fig.\ \ref{pillar})
consists of a pillar structure composed of a chain of asymmetric quantum 
dots separated by intervening layers of higher bandgap composition 
fabricated in a GaAs/AlGaAs technology by means of a sequence of planar 
MBE growth steps and subsequent etching to form the pillar. A sheath of 
similar AlGaAs composition is then grown surrounding the pillar and
a wrap-around gate electrode deposited. A drain (source)
is formed at the top (bottom), the series of asymmetric quantum dots 
are in the center region, and the gate surrounds the region of the pillar 
containing the quantum dots. Tarucha et al.\ \cite{bib:Tarucha96} have 
reported similar n-type single electron transistor (SET) structures. 
Electron confinement along the pillar axis is produced by the band gap 
discontinuity of the dot structure. Encasing the quantum dot structure 
in the pillar core by the cylindrical sheath and the gate electrode 
provides confinement in the radial direction. By applying a negative bias 
that depletes carriers near the surface, an additional parabolic 
electrostatic potential is formed that allows for tuning of the radial 
confinement and localization of one electron per dot. The simultaneous 
insertion of a single electron per dot is accomplished by lining up the 
quantum dot ground state levels so that they lie close to the Fermi level; 
a single electron is confined in each dot over a finite range of the gate 
voltage due to shell filling effects. \cite{bib:Tarucha96}

Thus, the pillar consists of a vertical stack of coupled asymmetric 
GaAs/AlGaAs quantum dots of differing size and composition so that each 
dot possesses a distinct energy structure. 
Qubit registers, $\arrowvert 0 \rangle$ and $\arrowvert 1 \rangle$, 
are based on the ground and first excited state of the single electron 
within each quantum dot.  Overall, parameters of the structure can be 
chosen to produce a well-resolved spectrum of distinguishable qubits.
The asymmetric dots produce large built-in 
electrostatic dipole moments between the ground and first excited state, 
and electrons in adjacent dots are coupled through the electric 
dipole-dipole interaction, while coupling between non-adjacent dots is 
significantly weaker.  This produces the desired quantum computer consisting 
of a linear array of binary states (qubits) with pair wise pillar-axis 
coupling between adjacent qubits. \cite{bib:collins}
In addition to energy tuning, the asymmetry of each quantum dot can be 
designed so that dephasing due to electron-phonon scattering and spontaneous 
emission is minimized. The combination of strong dipole-dipole coupling and 
long dephasing times make it possible to perform many computational steps 
before loss of coherence, in fact, it is believed possible to design this 
device so that error correction substantially prohibits coherence loss. 

Quantum computations are performed by means of a series of coherent optical 
pulses in the far infrared. Final readout of the amplitude and phase of the 
qubit states can be achieved through quantum state holography. Amplitude 
and phase information are extracted through mixing the final state with 
a reference state generated in the same system by an additional delayed 
laser pulse and detecting the total time- and frequency-integrated 
fluorescence as a function of the delay. \cite{bib:Leichtle}
%
%
Extracting the final state information using quantum state holography requires
multiple experiments, one for each delay, as described in Ref.
\onlinecite{bib:Leichtle}. Thus, the computation must be performed several times
before an answer is arrived at. This is no real problem since the number of
repetitions needed is only on the order of 40 or so, independent of the 
number of computational steps in a given quantum algorithm.
%
Through the use of integrated circuit manufacturing technology, it is 
possible to simultaneously fabricate a large array of "identical" pillar 
quantum dot quantum computers, that is, on the order of $ 10^{10} $ per 
wafer.  Each of these quantum registers could be electrically 
connected through deposited interconnect in such a manner so that each 
could be individually tunable to produce an array of identical units.
%
In general, inhomogeneity among the quantum dots will result in slightly 
different energy levels. Sherwin et al. \cite{bib:Sherwin} have recently
pointed out that one can perform accurate qubit operations in an inhomogeneous
population of quantum dots arising from quenched disorder due to static
charged defects, for example, provided that each SET is independently 
calibrated. This calibration can done by performing simple gate operations
and tuning the gate electrodes appropriately.
%
Efficient optical coupling to the resulting ensemble can be achieved 
through optical light guiding as suggested in Ref.\ \onlinecite{bib:mekis}. 
By this means direct observation of fluorescence is possible. Quantum 
computations are performed by means of a series of coherent optical pulses 
in the far infrared, and may be carried out in complete analogy with the 
operation of an NMR quantum computer. \cite{bib:cory}
%
%
It should be remarked that while our scheme resembles NMR ensemble quantum
computation in the use of a series of optical pulses to perform quantum
logic gates, it differs from NMR quantum computation in that our use of a
collection of single electron transistors is done to enable a stronger signal
to noise ratio in the readout phase. In principle, the quantum computation
could be done with only a single SET transistor structure if the readout
measurements were sufficiently sensitive.

\section{Device Model}

In the context of studies of the Coulomb blockade in self-organized
quantum dots and planar single-electron transistors,
self-consistent calculations of electronic structure, shell filling effects,
electron-electron interaction, Coulomb degeneracy, and
Coulomb oscillation amplitudes have been carried out for various
quantum dot structures.
\cite{bib:Averin91,bib:Stopa93,bib:Macucci93,bib:Jovanovic94,bib:Stopa96,bib:Todorovic97,bib:Macucci97,bib:Nagaraja97,bib:Fonseca98}
Our quantum register can be analyzed using methods similar to those
used to study the self-consistent electronic structure in
single-electron transistors. The problem we address is similar to
those addressed by other authors who are interested in obtaining 
current-voltage characteristics and studying Coulomb oscillations in
single-electron transistors over a wide range of gate biasing and shell
filling conditions.
\cite{bib:Averin91,bib:Stopa93,bib:Macucci93,bib:Jovanovic94,bib:Stopa96,bib:Todorovic97,bib:Macucci97,bib:Nagaraja97,bib:Fonseca98}

In our case, we are
interested in obtaining the self-consistent electrostatic potential
and electronic eigenstates in an equilibrium configuration in which
the source and drain are grounded and the gate electrode is negatively
biased. The electrostatic potential, $V(r)$, is obtained by solving the
Poisson equation for n-doped semiconductors
\cite{bib:Jovanovic94,bib:Fonseca98}
\begin{equation}
\nabla^{2} V(\vec{r}) = -\frac{4 \pi}{\varepsilon} \ q
\left[ n(\vec{r})+N_{D}^{+}(\vec{r}) \right]
\end{equation}
In the Poisson equation, $q$ is the absolute value of the electron charge,
$\varepsilon$ is the static dielectric constant, $n(\vec{r})$ is the electron
concentration and $N_{D}^{+}(\vec{r})$ is the known concentration of ionized
donors in the structure. For the dielectric constant, we adopt the GaAs value
$\varepsilon = 12$. \cite{bib:Pankove} The Poisson equation is solved subject
to boundary conditions on the electrostatic potential, $V(\vec{r})$. At the
interfaces between the semiconductor and the
source, drain and gate electrodes, $V(\vec{r})$ is equal to the applied gate
voltage while at the semiconductor-vacuum interfaces, the normal derivative
of $V(\vec{r})$ vanishes.

Following Ref. \onlinecite{bib:Jovanovic94},
the global electron concentration, $n(\vec{r})$, in the device is obtained by
partitioning the pillar structure into "bulk" and "quantum" regions. In
the "bulk" regions far from the quantum wells i.e. the 
source and drain regions, electrons are treated
in the Thomas-Fermi approximation and the electron concentration is
given by \cite{bib:Bethe}
\begin{equation}
n(\vec{r}) = \left\{
       \begin{array}{ll}
     \frac{1}{3 \pi^{2}}
     \left[ \frac{2m_{e}^{*}}{\hbar^2}\ (\mu-U(\vec{r})) \right]^{3/2}
           & \mbox{\ if \ $U(\vec{r}) < \mu$} \\
         0 & \mbox{\ otherwise}
        \end{array}
        \right.
\end{equation}
where $\mu$ is the chemical potential and $U(\vec{r})$ is the effective 
electron potential.
The chemical potential, $\mu$, is determined through the requirement
that overall charge neutrality be maintained in the bulk regions, i.e.
the chemical potential is adjusted until
\begin{equation}
\int \left( \ n(\vec{r}) - N_{D}^{+}(\vec{r}) \ \right) \ d\vec{r} = 0
\end{equation}
where the integration is carried out over the bulk source and drain
regions.

The effective potential, $U(\vec{r})$,
in the bulk regions includes the Hartree potential, $U_{H}=-q \ V(\vec{r})$,
and the conduction band offset, $\Delta E_{c}$, which depends on the
local Al concentration, $x$. Thus,
\begin{equation}
U(\vec{r}) = -q \ V(\vec{r}) + \Delta E_{c}
\end{equation}
where the conduction band offset, $\Delta E_{c}$, is taken to be 60\%
of the difference between the $Al_{x}Ga_{1-x}As$ and $GaAs$ bandgaps.
Using the bandgap variation of $Al_{x}Ga_{1-x}As$ determined by
Lee et al., \cite{bib:Lee}  we obtain the following expression for
the conduction band offset as a function of the local Al
concentration, $x$:
\begin{equation}
\Delta E_{c}= 0.6 \ (1155 \ x +370 \ x^{2}) \ \text{meV}
\end{equation}

In the "quantum" regions containing the quantum wells, the
electron concentration, $n(r)$, is determined by the electron
wavefunctions, $\psi_{i}(r)$, and energies, $E_{i}$, through the relation
\begin{equation}
n(\vec{r}) = \sum_{i} n_{i} \ \arrowvert \psi_{i}(\vec{r}) \arrowvert^{2}
\end{equation}
The electron occupancy in each level, $n_{i}$, is a function of
the electron energy and the temperature.

The electron wavefunctions and energy levels, $E_{i}$, are obtained by
solving the  Schr\"{o}dinger equation in the effective mass approximation
\begin{equation}
\left[-\frac{\hbar^{2}}{2 \ m_{e}^{*}}
\nabla^{2}+U(\vec{r})-E_{i} \right] \ \psi_{i}(\vec{r}) = 0 
\end{equation}
The electron potential, $U(\vec{r})$, in the quantum regions is given by
\begin{equation}
U(\vec{r}) = -q \ V(\vec{r}) + \Delta E_{c} + U_{xc}(\vec{r}) 
\end{equation}
where $U_{xc}(\vec{r})$ is the exchange-correlation potential of Perdew and
Zunger. \cite{bib:PerdewZunger}

In the quantum register discussed in the next section, the gate voltage
is negatively biased in such a way that a single electron is strongly
localized in each electrostatically confined quantum dot. The radial
confinement potential is strong enough that the lowest
few electron wavefunctions are strongly localized near the center of
the pillar and die away far from the semiconductor-electrode interface.
In our design, the quantum wells are wide enough and the barriers between
the quantum wells are thick enough so that the lowest
few electron wavefunctions do not penetrate to the center of the 
barriers separating the quantum wells. Since all the wavefunctions of
interest vanish at the center of these barriers, we can divide the
quantum well region into several regions (one for each qubit).
These regions are taken to be cylinders stacked 
along the pillar axis with top and bottom surfaces located at the
centers of the barriers between adjacent wells. We solve the
Schr\"{o}dinger equation in each dot separtely subject to the boundary
condition that all wavefunctions vanish at the region boundaries.

Due to the cylindrical symmetry of the structure, the 3D Schr\"{o}dinger
equation can be reduced to a 2D equation in cylindrical coordinates.
One might try to solve the 2D Schr\"{o}dinger equation by finite differencing
the partial differential equation and solving the resulting matrix eigenvalue
equation. The size of the matrix to be diagonalized is equal to the
number of interior mesh points in the 2D grid and this is much too large to
be handled easily. Other authors have taken this brute-force approach to
solving the Schr\"{o}dinger equation in self-consistent
Poisson-Schr\"{o}dinger problems with the result that solving the
Schr\"{o}dinger equation is the most time consuming part of the computation.
\cite{bib:Jovanovic94}  We find that it is possible to do better.
In solving the 2D Schr\"{o}dinger equation, we first approximate $U(\rho,z)$
in each quantum dot by a separable potential
\begin{equation}
U(\rho,z) \approx U_{s}(\rho,z) \equiv U_{r}(\rho) + U_{z}(z)
\end{equation}
where the axial potential is defined as
\begin{equation}
U_{z}(z) = \frac{2}{R^{2}} \int_{0}^{R} U(\rho,z) \ \rho \ d\rho
\end{equation}
and the radial potential is given by
\begin{equation}
U_{r}(\rho) = \frac{1}{\cal L} \int_{0}^{\cal L} 
\left( \ U(\rho,z) - U_{z}(z) \ \right) \ dz
\end{equation}
In these last two expressions, $R$ and ${\cal L}$ are the radius and height
of the cylindrical region over which $U(\rho,z)$ is defined in each dot.
With the separable potential approximation, the 2D Schr\"{o}dinger
equation can be separated into two 1D equations which can be cast
as finite difference eigenvalue equations and solved numerically
for the electron energies and wavefunctions. The resulting
2D wavefunctions are the best product wavefunctions that approximate
the solution of the 2D Schr\"{o}dinger equation in each qubit.
The electronic states in the separable
potential approximation in our cylindrical pillar are labeled by three
quantum numbers $(n_{\rho}$, $n_{\phi}$, $n_{z})$ which specify the
number of nodes in the product wavefunctions associated with
cylindrical coordinates $\rho$, $\phi$, and $z$. In this notation,
the qubit state $\arrowvert 0 \rangle$ is denoted $(0,0,0)$ while
$\arrowvert 1 \rangle$ is denoted $(0,0,1)$.
We find that the separable wavefunctions are reasonable approximations
to the true wavefunctions since we are starting with a separable potential
which is already close to the true potential in some average sense. 
We next obtain the exact energies and wavefunctions of the original
non-separable Schr\"{o}dinger equation by treating the residual
$U(\rho,z)-U_{s}(\rho,z)$ as a perturbation and expanding the exact
wavefunctions as a sum of separable wavefunctions.
Our expansion of the true wavefunctions in terms of separable wavefunctions
is rapidly converging and we find that the dominant terms in the expansion
of the true wavefunctions are the separable wavefunctions of the same
symmetry. Our approach to
solving the 2D Schr\"{o}dinger equation is fast and most
of computing time is spent solving the Poisson equation.

To complete the specification of the electron charge density in the
quantum dots, it is necessary to compute the electron occupation
numbers, $n_{i}$. One might expect that $n_{i}$ would be
given by the Fermi-Dirac distribution and indeed this would
be the case if the electrons in the dots were delocalized and
in tunneling contact with the leads. In this case,
the qubits could exchange electrons with their environment and
the total number of electrons in the dot $N = \sum_{i} n_{i}$ 
could take on non-integer values. But clearly this is
not tolerable in a quantum computer and we must carefully arrange
things so the dot wavefunctions exhibit a high degree of
localization. In this situation, only an integer number of electrons
can occupy the dot and this constraint gives rise to what is known
as the Gibbs distribution. The number of electrons, $N$, in the 
dot is determined by minimizing the Gibbs free energy with
respect to the integer number of electrons, $N$. The Gibbs free energy
is $F(N) = -kT \ \ln[Z(N)]$,
where the grand canonical partition function, $Z(N)$, is given by
\cite{bib:Jovanovic94,bib:Nagaraja97}
\begin{equation}
Z(N) = \sum_{\{n_{i}\}} \exp \left[
\frac{\sum_{i} n_{i}E_{i} - E_{H}(N) - \mu N}
{kT} \right]
\end{equation}
The lack of diffusive contact between the
quantum dots and the rest of the device means that the chemical
potential, $\mu$, is determined by electrons in the leads and contacts.
The summation in $Z(N)$ is carried out over all electron configurations
$\{n_{i}\}$ for which $\sum_{i} n_{i} = N$. Double counting the
Coulombic interaction is avoided by subtracting the Hartree energy
$E_{H}(N)$ for the $N$ electrons. The Hartree energy appearing 
in the partition function is
\cite{bib:Nagaraja97}
\begin{equation}
E_{H}(N) = \frac{1}{8 \pi \varepsilon}
\int \frac{n_{e}(\vec{r}) \ n_{e}(\vec{r}')}
{\arrowvert \vec{r}-\vec{r}' \arrowvert} \ d\vec{r} \ d\vec{r}'
\end{equation}
where $n_{e}(\vec{r})$ is the charge in the quantum dot and the integration
is restricted to the dot region.
Directly solving for the Hartree energy by performing a double 
integral over the quantum dot charge density is too time
consuming and impractical due
to the presence of the singularity in the integrand at $\vec{r} = \vec{r}'$.
An alternative method of calculating the Hartree energy is to
use the equivalent expression
\begin{equation}
E_{H}(N) = \frac{1}{8 \pi \varepsilon}
\int V_{e}(\vec{r}) \ n_{e}(\vec{r}) \ d\vec{r} 
\end{equation}
where the potential $V_{e}(\vec{r})$ is obtained by solving the Poisson
equation in the pillar using the charge density, $n_{e}(\vec{r})$,
{\em in the quantum dot}. \cite{bib:Fonseca98}

The boundary condition on $V_{e}(\vec{r})$ at the surface of the pillar is
determined by asymptotically expanding $V_{e}(\vec{r})$ in a multipole
expansion in the quantum dot charge density up through quadrupole terms
and using this expansion to specify $V_{e}$ at the surface.
This is a good approximation since the pillar boundaries are far from the 
localized quantum dot charge. \cite{bib:Fonseca98} 

To obtain a self-consistent solution to the coupled Poisson and
Schr\"{o}dinger equations, we first specify the device structure
including the $Al_{x}Ga_{1-x}As$ alloy composition, the doping
concentration, and the arrangement of the electrodes. In all
our runs, the source and drain are assumed to be grounded and the
gate is assumed to be negatively biased. We initially assume complete
depletion in the structure and solve the Poisson equation to obtain an
initial guess for the electrostatic potential, $V(\rho,z)$.
With this electrostatic potential and the quantum well band offset
potentials, we solve the Schr\"{o}dinger equation for the unoccupied
quantum dot energies and wavefunctions. The chemical potential
in the depleted structure is set to the minimum of the Thomas-Fermi
electron potential, $U = -q \ V(\rho,z) + \Delta E_{c}$, in the source
and drain regions. Starting with these initial guesses for the
chemical potential in the leads and the solutions of the Poisson and
Schr\"{o}dinger equations, we obtain self-consistent solutions through
the following relaxation procedure.

First, electron densities in the leads and
the quantum dots are obtained from the chemical potential, the temperature,
and the quantum dot electronic states. The global charge density, including
the given doping charge, is then obtained. Next the Poisson equation
is solved for $V(\rho,z)$. The Hartree potential and
exchange-correlation potentials are then obtained from $V(\rho,z)$ and the
electron charge density. With the electron potentials in hand,
the Schr\"{o}odinger
equation is solved in each quantum dot region. The procedure is then
repeated until convergence is achieved. In updating the
electrostatic potential and electron charge density, the new solutions
are mixed with the old to obtain the updated solutions. For the
electrostatic potential
\begin{equation}
V(\rho,z) \rightarrow \lambda \ V_{new}(\rho,z) + (1-\lambda) \ V_{old}(\rho,z)
\end{equation}
where $\lambda < 1$ is a relaxation parameter which is
dynamically adjusted to accelerate convergence. A similar scheme is
used to update the electron charge density. The above procedure
is iterated until the chemical potential, electrostatic potential,
electron charge density, and quantum dot energy levels all change by
less than some small relative tolerance between successive iterations
at which point convergence is achieved. Typically about 400 iterations
are required to achieve convergence to within one part in $10^{4}$.

\section{A Three qubit quantum register: 1D analysis}

We can use the device modeling program described in the last section to
obtain a design for a three-qubit quantum register. We could, in principle,
do a full 3D analysis of the device and obtain suitable design parameters
(i.e., pillar dimensions, doping concentrations, asymmetric well shapes,
electrode placement and biasing, etc.) based on our computationally
intensive 3D model. Clearly this would be prohibitively time consuming due
to the size of the parameter space that would need to be investigated as
well as the time required to perform each run. To narrow down the design
parameters, we can take advantage of the fact that our quantum computer
is operated in the extreme depletion limit and do a simple 1D analysis
to gain some useful insight.

Let's assume that inside the core of stacked quantum wells (radius $R_{c}$)
we have {\em complete} depletion and uniform doping. In this limit,
the quantum dot electron potential, $U(r)$, can be expressed in
cylindrical coordinates as $U(\vec{r}) = U(z) + U(\rho)$, where
$U(\rho)$ is a radial potential arising from the uniform donor density 
and $U(z)$ is the conduction band offset potential along the growth direction.
This separable potential assumption is a good approximation in the strong
depletion regime where only a single electron resides in each dot.
The assumption of a separable potential is commonly used in the study
of quantum dot structures and enables us to consider the $z$ and
$\rho$ motions separately. \cite{bib:Tarucha96,bib:Jacak98}  The z-directional 
potential $U(z)$, shown schematically in the inset of Fig.\ \ref{density},
is a step potential formed by a layer of $Al_{x}Ga_{1-x}As$ of thickness 
$B$ ($0 < z < B$) and a layer of $GaAs$ of thickness $L - B$ ($B < z < L$).
The resulting asymmetric quantum dot/well is confined by $Al_{y}Ga_{1-y}As$
barriers with $y > x$ and the asymmetry is parameterized by the ratio $B/L$
where $0 < B/L < 1$.
    
In the effective mass approximation, the qubit wavefunctions are
$\arrowvert i \rangle = R(\rho) \ \psi_{i}(z) \ u_{s}(\vec{r})$ ($i = 0,1$).
Here $R(\rho)$ is the ground state of the radial envelope function,
$\psi_{i}(z)$ is the envelope function along $z$,
and $u_{s}(\vec{r})$ is the $s$-like zone center Bloch function including
electron spin. For simplicity, we assume complete confinement by the
$Al_{y}Ga_{1-y}As$ barriers along the z direction. Then, the envelope 
function $\psi_{i}(z)$ is obtained by solving the time-independent 
Schr\"{o}dinger equation subject to the boundary conditions
$\psi_{i}(0) = \psi_{i}(L) = 0$. The energies of the qubit wavefunctions
are given by $E = E_{\rho} + E_{i}$ where $E_{\rho}$ is the energy
associated with $R(\rho)$ and $E_{i}$ is the energy associated with
$\psi_{i}(z)$.

Figure \ref{density} shows the probability density,
$\arrowvert \psi_{i}(z) \arrowvert ^{2}$, as a function of position, 
$z$, for the two qubit states $\arrowvert 0 \rangle$ and
$\arrowvert 1 \rangle$ in a $20 \ nm$ $GaAs/Al_{0.3}Ga_{0.7}As$
asymmetric quantum dot. The barrier thickness $B = 15 \ nm$ and the 
overall length of the dot is $L = 20 \ nm$. By choosing $B/L = 0.75$ and
$x = 0.3$, it is found that the ground state wavefunction
$\arrowvert 0 \rangle$ is strongly localized in the $GaAs$ region while the
$\arrowvert 1 \rangle$ wavefunction is strongly localized in the
$Al_{0.3}Ga_{0.7}As$ barrier. By appropriately choosing the
asymmetric quantum dot parameters, the qubit wavefunctions can be
spatially separated and a large difference in the electrostatic dipole
moments can be achieved.

The transition energy $\Delta E_{0} = E_{1} - E_{0}$ between
$\arrowvert 1 \rangle$ and $\arrowvert 0 \rangle$ is shown in 
Fig.\ \ref{transition} as a function of $B/L$ in a $20 \ nm$
$GaAs/Al_{x}Ga_{1-x}As$ asymmetric quantum dot ($ L= 20 \ nm$).
Several values of Al concentration $x$ are considered.
In Fig.\ \ref{qubit_energy}, we fix the Al
concentration at $x = 0.2$ and plot $\Delta E_{0}$ as a function of
$L$ for several values of $B/L$. The continuous
curves are based on our 1D analysis and the squares are the
qubit energy gaps for a three qubit self-consistent quantum register
calculation as described in the next section. It is clear that
the transition energy can be tailored substantially by varying the
asymmetry parameter. With three parameters available for adjustment
($B$, $L$, and $x$), we can make $\Delta E_{0}$ unique for each dot in the
register. In this way, we can address a given dot by using laser light
with the correct photon energy.

It is desirable that the $\arrowvert 1 \rangle$ state be the first excited
level of the quantum dot. Thus, the lowest lying radial state $(0,1,0)$
should lie above the $\arrowvert 1 \rangle$ state. The radial energy
gap, $\Delta E_{1}$, between the ground state, $\arrowvert 0 \rangle$,
and the first radial excited state, $(0,1,0)$, is found by solving
a 2D Schr\"{o}dinger equation for an electron in the radial potential,
$U_{r}(\rho)$. If we take the barrier in the sheath to be infinite, then
in the extreme depletion limit, we have
\begin{equation}
U_{r}(\rho) = \left\{
       \begin{array}{ll}
         -q \ V(\rho)      & \mbox{\ if $\rho < R_{c}$} \\
         \infty            & \mbox{\ otherwise}
        \end{array}
        \right.
\end{equation}
where $V(\rho)$ is the radial electrostatic potential.
For complete depletion and uniform doping, the Poisson equation for
$V(\rho)$ can be solved analytically. Thus,
\begin{equation}
V(\rho) = \frac{\pi N_{D}^{+}}{\varepsilon} (R_{c}^2 - \rho^2)
\end{equation}
where $R_{c}$ is the sheath radius and $N_{D}^{+}$ is the doping density.
Numerically solving the 2D Schr\"{o}dinger equation for an electron in
the potential, $U_{r}(\rho)$, is straightforward.
Figure \ref{radialgap} shows the radial energy gap, $\Delta E_{1}$,
between the $\arrowvert 0 \rangle$ and the lowest lying radial
state, $(0,1,0)$, as a function of doping concentration, $N_{D}^{+}$,
for several values of $R_{c}$. For narrow pillars with low doping
concentrations, the radial energy gap is determined by size confinement.
For large pillars with high doping concentrations the radial energy gap
is determined by electrostatic confinement.  From Fig.\ \ref{qubit_energy}, 
we see that the qubit energy gaps reach a
minimum near $\Delta E_{0} \sim 70 \ meV$ for quantum wells
with $L \sim 20 \ nm$.  The results of Fig.\ \ref{radialgap} suggest that
radial gaps in the range of $\Delta E_{1} \sim 100 \ meV$ with strong
size confinement can be achieved with doping densities in the
range of $N_{D}^{+} \sim 10^{17} \ cm^{-3}$ if $R_{c} \sim 70 \ \AA$. 

The electric field from an electron in one dot shifts the energy
levels of electrons in adjacent dots through electrostatic dipole-dipole
coupling. By appropriate choice of coordinate systems, the dipole moments
associated with $\arrowvert 0 \rangle$ and $\arrowvert 1 \rangle$
equal in magnitude but oppositely directed. The dipole-dipole
coupling energy is then defined as \cite{bib:BarencoDeutsch95}
\begin{equation}
V_{dd}=2 \ \frac{\arrowvert d_{1} \arrowvert \ \arrowvert d_{2} \arrowvert}
{\epsilon_{r} R_{12}^{3}},
\label{Vdd}
\end{equation}
where $d_{1}$ and $d_{2}$ are the ground state dipole moments in the
two dots, $\epsilon_{r} = 12.9$ is the dielectric constant for $GaAs$,
and $R_{12}$ is the distance between the dots.

Figure \ref{coupling} shows the dipole-dipole coupling energy, $V_{dd}$, 
between two asymmetric $GaAs/Al_{x}Ga_{1-x}As$ quantum dots of 
widths $L1 = 19 \ nm$
and $L2 = 21 \ nm$ separated by a $10\ nm$ $Al_{y}Ga_{1-y}As$ barrier.
The coupling energy is plotted as a function of $B/L$ for several values of
$x$ where $B/L$ and $x$ are taken to be the same in both dots. The
dipole-dipole coupling energies are a strongly peaked
function of the asymmetry parameter, $B/L$. From the figure, we see that
values of $V_{dd} \sim 0.15 \ meV$ can be achieved.

Quantum dot electrons can interact with the environment through the
phonon field, particularly the longitudinal-optical (LO) and acoustic (LA) 
phonons. The LO phonon energy, $\hbar \omega_{LO}$, lies in a narrow
band around $36.2 \ meV$. As long as the quantum
dot energy level spacings lie outside this band, LO phonon scattering
is strongly suppressed by the phonon bottleneck effect. Acoustic
phonon energies are much smaller than the energy difference, $\Delta E$,
between qubit states. Thus, acoustic phonon scattering requires
multiple emission processes which are also very slow. Theoretical
studies on phonon bottleneck effects in GaAs quantum dots indicate that
LO and LA phonon scattering rates including multiple phonon processes
could be slower than the spontaneous
emission rate {\em provided that the quantum dot energy level spacing is
greater than $\sim 1\ meV$ and, at the same time, avoids a narrow window 
around the LO phonon energy}. \cite{bib:Inoshita,bib:Bockelman,bib:Benisty} 
In Ref.\ \onlinecite{bib:Inoshita}, Inoshita and Sakaki
compute multi-phonon relaxation rates in spherical single-electron $GaAs$
quantum dots due to one- and two-phonon scattering by LO and LA
phonons at $T = 0 \ K$ and $T = 300 \ K$. Using the results of this
calculation, we estimate that multi-phonon scattering dominates the
spontaneous emission only if the qubit energy level spacing is
within $\sim 4 \ meV$ of the LO phonon energy. Likewise, multi-phonon
LA scattering becomes important if the qubit energy gaps are smaller
than $\sim 1 \ meV$ 

While dephasing via interactions with the phonon field can be strongly 
suppressed by proper designing of the structure, quantum dot electrons
are still coupled to the environment through spontaneous emission and
this is the dominant dephasing mechanism. Decoherence resulting from
spontaneous emission ultimately limits the total time available for a
quantum computation. \cite{bib:Ekert96}  Thus, it is important that the
spontaneous emission lifetime be large. The excited state lifetime,
$T_{d}$, against spontaneous emission is \cite{bib:Ekert96}
\begin{equation}
T_{d} = \frac{3 \hbar \ (\hbar c)^{3}}{4 e^{2}\ D^{2}\ \Delta E^{3}} ~ ,
\label{Td}
\end{equation}
where $D = \langle 0 \arrowvert z \arrowvert 1 \rangle$
is the dipole matrix element between $\arrowvert 0 \rangle$
and $\arrowvert 1 \rangle$.

Figure \ref{lifetime} shows the spontaneous emission lifetime of an electron
in qubit state $\arrowvert 1 \rangle$ for a $20 \ nm$ $GaAs/Al_{x}Ga_{1-x}As$
quantum dot as a function of asymmetry parameter, $B/L$, for several
values of Al concentration, $x$. It is immediately obvious from
Fig.\ \ref{lifetime} that the lifetime depends strongly on $B/L$. Depending
on the value of $x$ chosen, the computed lifetime can achieve a maximum
of between 4000 $ns $ and 6000 $ns$.
In general, the maximum lifetime increases with $x$. In Eq.\ (\ref{Td}),
the lifetime is inversely proportional to $\Delta E^{3}$ and $D^{2}$, but
the sharp peak seen in Fig.\ \ref{lifetime} is due {\em primarily}
to a pronounced minimum in $D$. 

Based on these results, we can estimate parameters for a solid state
quantum register containing a stack of several asymmetric
$GaAs/Al_{0.3}Ga_{0.7}As$ quantum dots in the $L \sim 20 \ nm$ range
separated by $10 \ nm$ $Al_{y}Ga_{1-y}As$ barriers ($y > 0.4$).
An important design goal is obtaining a large
spontaneous emission lifetime and a large dipole-dipole coupling 
energy.  From Figs. \ref{coupling} and \ref{lifetime}, we see that both can
be achieved by selecting an asymmetry parameter, $B/L = 0.8$.
This gives us a spontaneous emission lifetime $T_{d} = 3100 \ ns$ 
and a dipole-dipole coupling energy $V_{dd} = 0.14 \ meV$.
The transition energy between the qubit states is on the order of
$100\ meV$ ($\lambda = 12.4 \ \mu m$). In a quantum computation, the
quantum register is optically driven by a laser as described in
Ref. \onlinecite{bib:BarencoDeutsch95}. In our example, we require a
tunable infra-red laser in the mid-$10\ \mu m$ range so we can individually
address various transitions between coupled qubit states.

\section{A Three qubit quantum register: 3D analysis}

Using the results of our simple 1D model as a starting point, we designed
a three qubit quantum register by using the self-consistent
device model described in Section III. Several criteria have to be
met for a viable quantum register design and the structure we
obtained through trial-and-error involved tradeoffs between
several design goals.

For a self-consistent quantum register calculation, we assume the
parameters of the free-standing quantum dot pillar structure (shown in 
Fig.\ \ref{pillar}) as follows:  The height of the pillar 
is taken to be $ {\cal L} = 1000 \ nm$ while the radii of the core and sheath
are taken to be $R_{c} = 7 \ nm$ and $R = 50 \ nm$.
The drain and source contacts at the top and bottom
of the pillar are grounded and a cylindrical gate with a height of
$400 \ nm$ is placed around the center of the pillar. Near the source and
drain contacts, layers of intrinsic semiconductor serve to inhibit
gate-to-source and gate-to-drain currents. The central $600 \ nm$ of the
pillar is uniformly n-doped with a doping concentration of
$N_{D} = 5 \times 10^{17} \ cm^{-3}$.

The cylindrical sheath surrounding the core region is composed
of high band  gap $Al_{0.45}Ga_{0.55}As$ and serves to confine electrons to
the core region.
The three qubits in the core are defined by the composition profile of
$Al_{x}Ga_{1-x}As$ along the pillar axis. In our structure, the
Al concentration, $x$, in the core region is uniform in the radial
direction. The composition profile along the pillar axis in the core region
is shown in Fig.\ \ref{composition}. The ground and first excited electronic
states are the qubit states $\arrowvert 0 \rangle$ and
$\arrowvert 1 \rangle$ and the electron charge densities for these
states are shown schematically in the figure. We find that in thermal
equilibrium the electrons reside entirely in the ground state
$\arrowvert 0 \rangle$ for temperatures as high as $77 \ K$
since the energy gap between $\arrowvert 0 \rangle$
and $\arrowvert 1 \rangle$ is much greater than $kT$.
This is indicated schematically by the solid circles
in the diagram. The widths $L$ of the asymmetric quantum wells/dots
defining qubits $1$ through $3$ are $19.0$, $20.5$ and $22.0 \ nm$
while the $B/L$ ratios are $0.670$, $0.683$ and $0.675$ respectively. 
Our 1D analysis suggests that asymmetry parameters in this range
will result in long spontaneous emission lifetimes and strong
dipole-dipole coupling between neighboring qubits. The asymmetric
quantum wells are composed of $GaAs$ and $Al_{0.2}Ga_{0.8}As$ layers
and the barriers between the asymmetric dots/wells are composed of
$Al_{0.45}Ga_{0.55}As$.

With a properly chosen reverse gate bias, $V_{g}$, the doping region
in the center of the pillar is depleted and the equilibrium 
Fermi level lines up so that there is exactly one electron in each dot.
Single electron occupancy in the dots is necessary in order for there to
be a {\em well defined qubit Hilbert space}. Due to shell filling
effects, single electron occupancy in all three dots holds over a finite
range of the gate voltage. By running our device model for several values
of $V_{g}$, we find that single electron occupancy is obtained
over the range $-1.56 ~V ~ \leq V_{g} \leq ~ - 1.48 ~ V$. Thus, the requirement
for single electron occupancy in the quantum dots is {\em maintained
in the presence of gate voltage fluctuations} on the order of
$\Delta V_{g} \approx 0.08 \ V$. For $V_{g} = -1.5 \ V$, the
self-consistent electron potential along the pillar axis,
(i.e. $\rho = 0$) is shown in Fig.\ \ref{electron_potential} as a
function of position along the pillar axis. The position along the
pillar axis is measured from the drain contact at $z = 0 \ nm$ to the
source contact at $z = 1000 \ nm$. Figure \ref{electron_potential}
is centered on the active region of the register containing the 
three quantum dots and the origin of the energy scale is chosen to be
the equilibrium Fermi level. The total electron potential is approximately
the sum of the self-consistent electrostatic Hartree potential and
the conduction band offset potential, the self-consistent
exchange-correlation potential being negligible.

The self-consistent electron levels are obtained by solving the
Schr\"{o}dinger equation in the self-consistent potential shown in
Fig.\ \ref{electron_potential}. In our structure, the
$\arrowvert 0 \rangle$ ground states have
$(n_{\rho},n_{\phi},n_{z}) = (0,0,0)$ symmetry and the
$\arrowvert 1 \rangle$ states (the first excited level) in all three
qubits are $(n_{\rho},n_{\phi},n_{z}) = (0,0,1)$ states.
The self-consistent qubit energy gap, $\Delta E_{0}$, between the
$\arrowvert 0 \rangle$ and $\arrowvert 1 \rangle$ states,
the radial energy gap, $\Delta E_{1}$, and
the spontaneous emission lifetime of the $\arrowvert 1 \rangle$ state,
$\tau_{s}$, and dipole moment, $d$, for the three qubits are listed
in Table \ref{qubit_levels}. From Table \ref{qubit_levels}, we
see that the radial energy gaps are larger than the qubit energy gaps.
Another thing to note is that the qubit energy gaps are large compared
to $k T$ at $T = 77 \ K$. Thus, in thermal equilibrium
the electrons reside entirely in the $\arrowvert 0 \rangle$ level at $77 \ K$.
This means that the initial state of the
quantum register is characterized by a pure state density matrix
$\hat{\rho}_{0} = \arrowvert 0,0,0 \rangle \ \langle 0,0,0 \arrowvert$.
Consequently, there is
{\em no need for initial state preparation} in our quantum register.
In Fig.\ \ref{electron_density}, the self-consistent electron probability
densities in the three quantum dots are plotted as a function of position
along the pillar axis. Each dot traps one electron and the probability
densities in the ground and first excited states are shown as solid and
dot-dashed lines, respectively. The barriers are thick enough so that
electron wavefunctions in adjacent dots do not overlap.

The energy levels for the three qubit quantum computer are shown in
Table \ref{multi-qubit} with and without the inclusion of dipole-dipole
coupling between the qubits. From Table \ref{multi-qubit} we see
that a different energy corresponds to each three-electron state
$\arrowvert i_{1}, \ i_{2}, \ i_{3} \rangle$ of the register where
$i_{n} = (0,\ 1)$ labels the state of the $n$-th qubit. Transition
energies between the states $\arrowvert 0 \rangle$ and
$\arrowvert 1 \rangle$ for a given qubit are obtained by taking
differences between the appropriate entries in Table \ref{multi-qubit}.
For the first qubit, we take differences between all three-particle
states $\arrowvert 0, \ i_{2}, \ i_{3} \rangle$ and
$\arrowvert 1, \ i_{2}, \ i_{3} \rangle$. In general, the transition
energy between $\arrowvert 0 \rangle$ and $\arrowvert 1 \rangle$
for an electron in the first qubit will depend on the
states, $i_{2}$ and $i_{3}$, occupied by the second and third qubits,
and there can be as many as four such 
conditional transitions. In the absence of dipole-dipole coupling
between qubits, all four conditional transition energies between 
$\arrowvert 0 \rangle$ and $\arrowvert 1 \rangle$ for a given
qubit are degenerate. When dipole-dipole interactions between the
qubits are considered, the four-fold degenerate conditional transition
energies split into multiplets depending on which states are occupied by
the electrons in neighboring qubits. 

The conditional transition energies
between $\arrowvert 0 \rangle$ and $\arrowvert 1 \rangle$ states
for our three qubit register are shown in Fig.\ \ref{spectrum} as a
function of photon energy. In the absence of dipole-dipole coupling, the
transition energies for the three qubits are $40.86 \ meV$, $47.14 \ meV$,
and $52.31 \ meV$, respectively. When dipole-dipole interactions between
qubits are taken into account, the conditional transition energies
split into multiplets as shown in this figure.  Each transition
in the spectrum is labeled by the neighboring electron states which give
rise to it. By performing optical $\pi$-pulses at selected conditional 
transition frequencies, quantum logic operations can be performed. For
example, a $\pi$ pulse performed on the lowest energy transition in
Fig.\ \ref{spectrum} performs a bit flip on the first qubit provided
the second qubit is in state $\arrowvert 1 \rangle$. This operation
is just a Controlled-Not gate with qubit 2 as the control bit and
qubit 1 as the target bit.

The need to selectively perform $\pi$-pulses at the conditional
transition frequencies allows us to make some preliminary estimates
on the parameters of the laser system needed to drive a quantum
computation. If we want to selectively drive a given transition
without exciting neighboring transitions, then the bandwidth of
the $\pi$-pulse needs to be less than the splitting between the
two most closely spaced lines in the spectrum.  From Fig.\ \ref{spectrum}, 
the two most closely spaced lines are
spaced $\Delta \hbar \omega \approx 0.0776 \ meV$ apart. If we require
that the $\pi$-pulse bandwidth is $\Delta E_{\pi} \approx 0.01 \ meV$,
then the pulse length can be estimated from Heisenberg's uncertainty
principle, $\Delta E_{\pi} \Delta T_{\pi} \approx \hbar/2$, to be
$T_{\pi} \approx 33 \ ps$. If we assume a square $\pi$-pulse, the
magnitude of the optical electric field is given by \cite{bib:Mahler98}
\begin{equation}
E_{0} = \frac{\pi \hbar}{q \ d \ T_{\pi}}
\end{equation}
where $d$ is the optical dipole from Table \ref{qubit_levels}
and the average Poynting vector during the pulse is
\cite{bib:Lorraine70}
\begin{equation}
S_{av} = \frac{1}{2} c \epsilon_{0} E_{0}^{2}
\end{equation}
For $d \approx 10 \ \AA$ and $T_{\pi} \approx 33 \ ps$, we obtain
$E_{0} \approx 0.627 \ kV/cm$ and $S_{av} \approx 522 \ W/cm^{2}$.

\section{Summary}

In this paper, we have studied a solid state implementation of
quantum computing based on coupled quantum dots. 
Our quantum register consists of a free standing n-type
pillar with grounding leads at the top and bottom of the
structure. Asymmetric quantum wells confine electrons along
the pillar axis and a high bandgap $AlGaAs$ sheath wrapped around the
center of the pillar allows for confinement in the radial
direction. The ground and first excited electronic states of the
quantum dots act as qubit states $\arrowvert 0 \rangle$ and
$\arrowvert 1 \rangle$, respectively.
We have developed a 3D device model for a general SET structure
containing several quantum dots. We self-consistently solve
coupled Schr\"{o}dinger and Poisson equations for the device and
develop a design for a three qubit quantum register with asymmetric
quantum dots tailored for long dephasing time and large dipole-dipole
coupling between the dots. Our results indicate that a single gate
electrode can be used to localize a single electron in each of the
quantum dots. Adjacent dots are strongly coupled by electric
dipole-dipole interactions arising from the dot asymmetry thus enabling
rapid computation rates.

\acknowledgments
This study was supported, in part, by the Defense Advanced Research Project 
Agency and the Office of Naval Research.



\begin{table}
\caption{Self-consistent qubit energy gap, $\Delta E_{0}$,
radial energy gap, $\Delta E_{1}$, spontaneous
emission lifetime, $\tau_{s}$, and optical dipole moment, $d$, for a
three qubit quantum register.}
\label{qubit_levels}
\begin{tabular}{ccccc}
Qubit No. & $\Delta E_{0} \ (meV)$ & $\Delta E_{1} \ (meV)$ &
$\tau_{s} \ (ns)$ & $d \ (\AA)$ \\
\hline
 1        & 40.86  &  63.7      &    28000  &    11.7     \\
 2        & 47.14  &  63.2      &    19000  &    11.5     \\
 3        & 52.31  &  61.9      &    14000  &    11.4
\end{tabular}
\end{table}

\begin{table}
\caption{Register energies for a three qubit quantum register,
with and without dipole-dipole coupling interaction between qubits.}
\label{multi-qubit}
\begin{tabular}{cdd}
 Register state & Energy (meV)  & Energy (meV)                \\
                & uncoupled     & dipole-dipole coupled       \\
\hline
$\arrowvert  0 \ 0 \ 0 \rangle$  &   -70.161    &    -70.357  \\
$\arrowvert  0 \ 0 \ 1 \rangle$  &   -17.852    &    -17.833  \\
$\arrowvert  0 \ 1 \ 0 \rangle$  &   -23.017    &    -22.822  \\
$\arrowvert  0 \ 1 \ 1 \rangle$  &    29.292    &     29.273  \\
$\arrowvert  1 \ 0 \ 0 \rangle$  &   -29.292    &    -29.311  \\
$\arrowvert  1 \ 0 \ 1 \rangle$  &    23.017    &     23.213  \\
$\arrowvert  1 \ 1 \ 0 \rangle$  &    17.852    &     17.871  \\
$\arrowvert  1 \ 1 \ 1 \rangle$  &    70.161    &     69.966  \\
\end{tabular}
\end{table}


\begin{figure}
\caption{Schematic illustration of the proposed quantum dot pillar structure.
As an example, the structure is shown for a three qubit quantum register 
(i.e., with three quantum dots in the middle).}
\label{pillar}
\end{figure}

\begin{figure}
\caption{Probability density along the confinement direction, $z$, for the
qubit wavefunctions $\arrowvert 0 \rangle$ (solid line)
and $\arrowvert 1 \rangle$ (dot-dashed line).  The inset shows a schematic 
illustration of the conduction bandedge profile in the $ z $ direction.}
\label{density}
\end{figure}

\begin{figure}
\caption{Transition energy, $\Delta E$, between $\arrowvert 0 \rangle$
and $\arrowvert 1 \rangle$ in an $L = 20 \ nm$
$GaAs/Al_{x}Ga_{1-x}As$ asymmetric quantum dot as a
function of $B/L$ for several values of $x$.}
\label{transition}
\end{figure}

\begin{figure}
\caption{The transition energy between the $\arrowvert 0 \rangle$
and $\arrowvert 1 \rangle$ qubit states as a function of the
asymmetric quantum well width, $L$, for several values of the
asymmetry parameter $B/L$ and fixed barrier Al concentration
$x = 0.2$. The continuous curves are based on a simple 1D analysis
and the squares are the qubit energy gaps for quantum dots as 
determined by the self-consistent calculation described in the text.}
\label{qubit_energy}
\end{figure}

\begin{figure}
\caption{The energy gap between the $\arrowvert 0 \rangle$
 ground state and the lowest lying radial eigenstate
(with $(n_{\rho},n_{\phi},n_{z}) = (0,1,0)$ symmetry)
as a function of the n-doping concentration, $N_{D}^{+}$. The
radial energy splitting for several pillar radii.
Greater confinement of the radial wavefunction can be obtained by
decreasing the pillar radius (size confinement) or increasing
the doping concentration (electrostatic confinement).
In the limit of large pillar radii, the radial energy splitting is
determined by confinement in a parabolic electrostatic potential.}
\label{radialgap}
\end{figure}

\begin{figure}
\caption{Dipole-dipole interaction between two asymmetric
$GaAs/Al_{x}Ga_{1-x}As$ quantum dots of widths $L1 = 19 \ nm$
and $L2 = 21 \ nm$ separated by an $Al_{y}Ga_{1-y}As$ barrier of width
$Wb = 10 \ nm$. The coupling energy is plotted as a function of $B/L$
for several values of $x$. $B/L$ and $x$ are the same for both dots.}
\label{coupling}
\end{figure}

\begin{figure}
\caption{Spontaneous emission lifetime for qubit state
$ \arrowvert 1 \rangle $ in a $GaAs/Al_{x}Ga_{1-x}As$ quantum dot 
with $ L = 20 \ nm$ as a function of $B/L$ for several values of $x$.}
\label{lifetime}
\end{figure}

\begin{figure}
\caption{Composition profile along the pillar axis for a three qubit
quantum register.}
\label{composition}
\end{figure}

\begin{figure}
\caption{Self consistent electron potentials as a function
of position along the pillar axis in the active region of a three qubit 
quantum register. The total potential (solid line) is the sum of the 
Hartree potential (dot-dashed line) and the band offset potential.
The fermi level (dotted line) is aligned so that exactly one electron
resides in each quantum dot.}
\label{electron_potential}
\end{figure}

\begin{figure}
\caption{Electron density as a function of position along the pillar
axis in the active region of a three qubit quantum register. The bias
voltage is adjusted so that exactly one electron resides in the 
ground state of each quantum dot. In each dot, the probability 
density of the ground states (solid line) are shown along with the
probability density of the first excited states (dot-dashed line).
Depending on which state the electrons occupy, the electric dipole
moment can point either left or right.}
\label{electron_density}
\end{figure}

\begin{figure}
\caption{Conditional transition energies between
the qubit states $\arrowvert 0 \rangle$ and $\arrowvert 1 \rangle$
as a function of photon energy for a three qubit quantum register.
In the absence of dipole-dipole coupling, the transition energies
are $40.86 \ meV$, $47.14 \ meV$, and $52.31 \ meV$ respectively. When
dipole-dipole interactions between qubits are taken into account, 
the transition energies depend on the state of the adjacent qubits.}
\label{spectrum}
\end{figure}

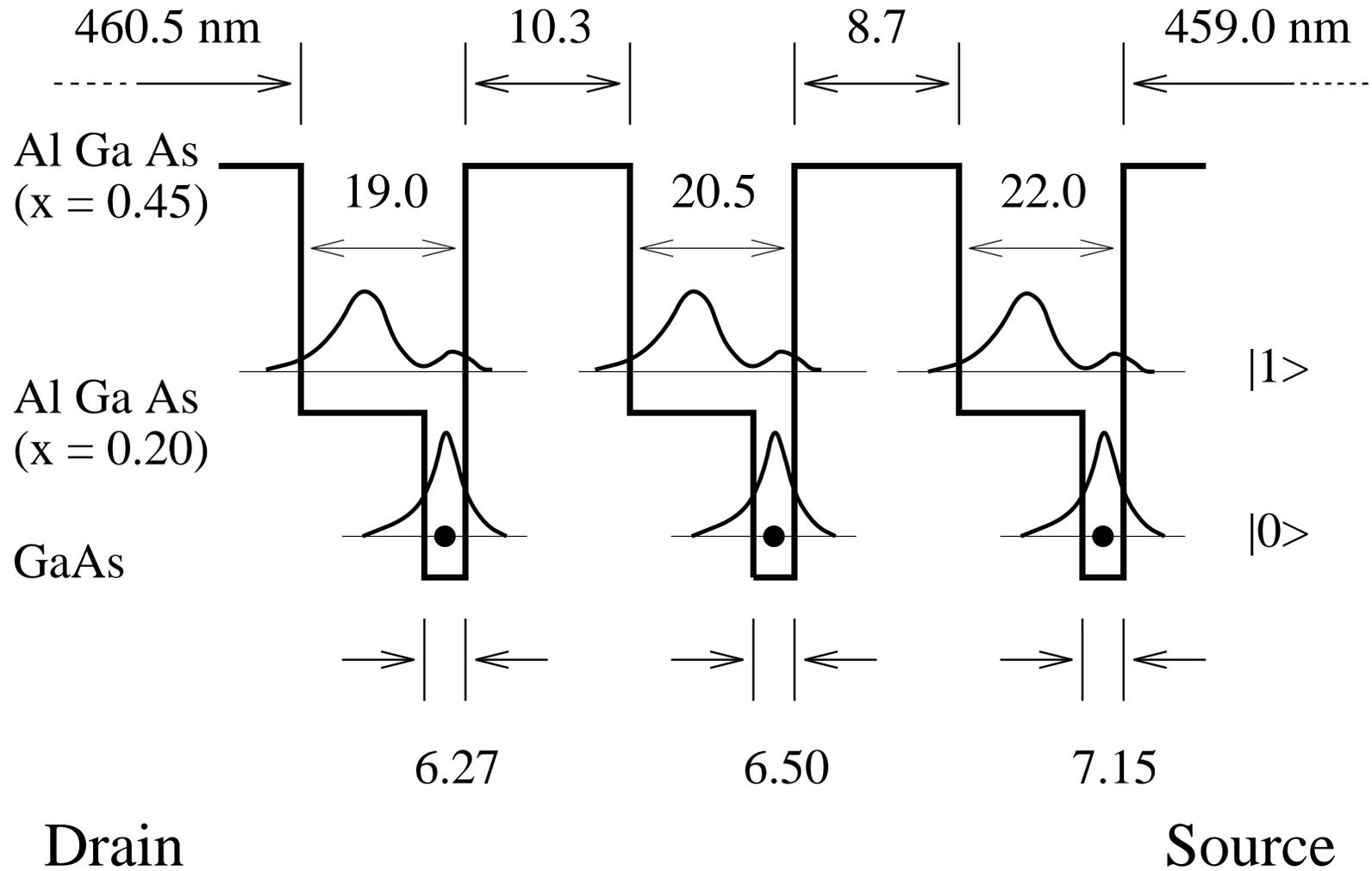
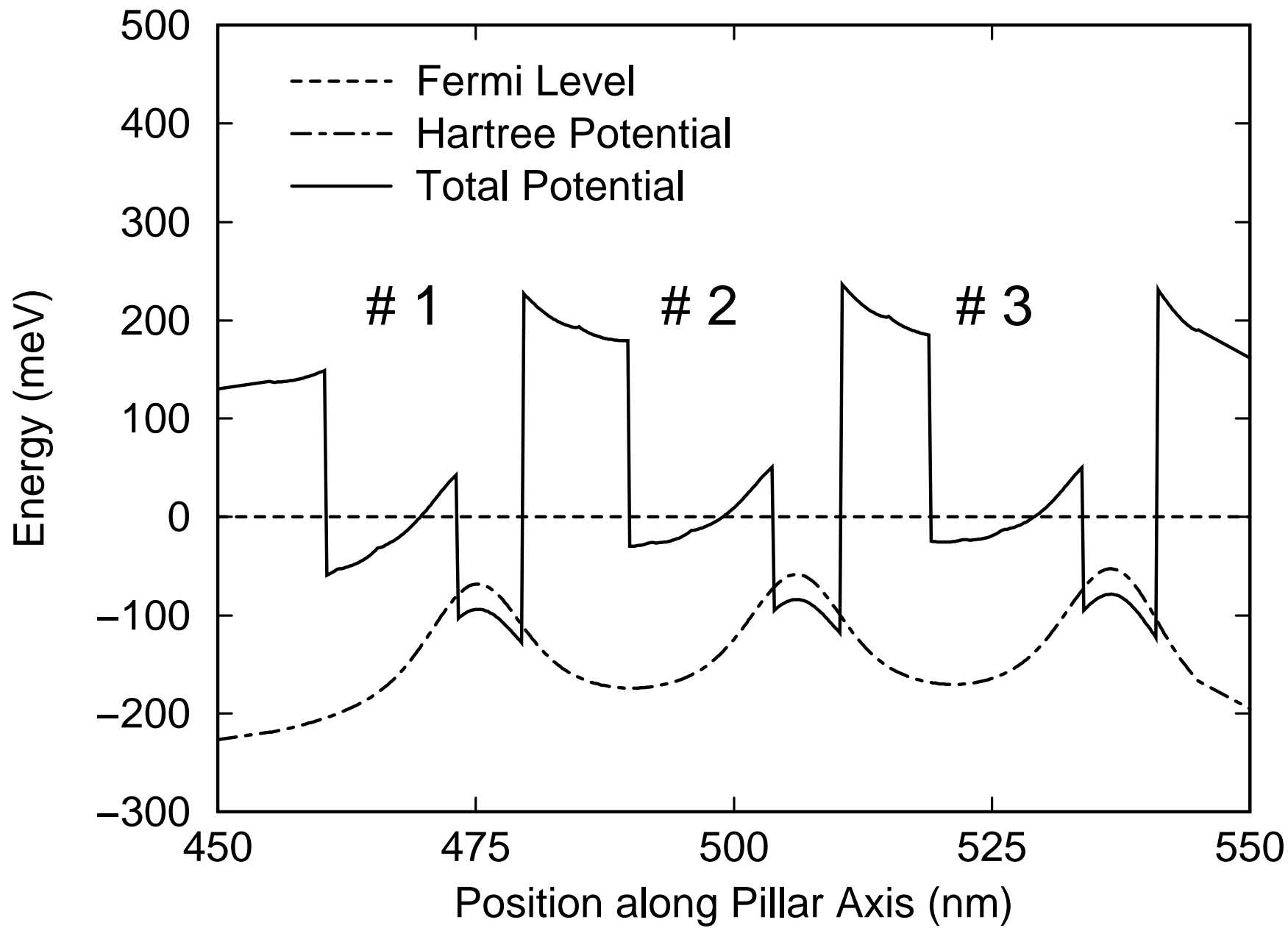
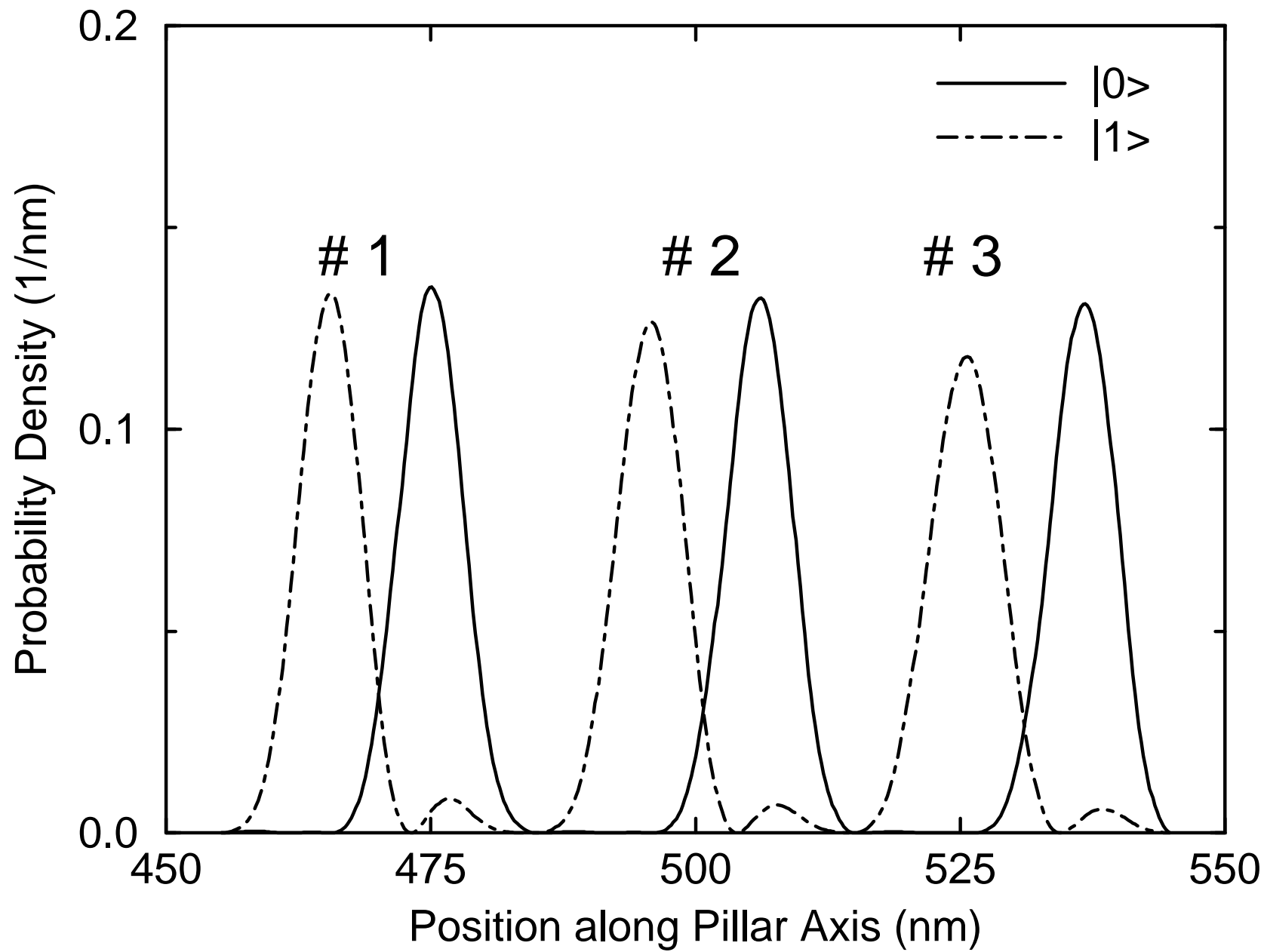
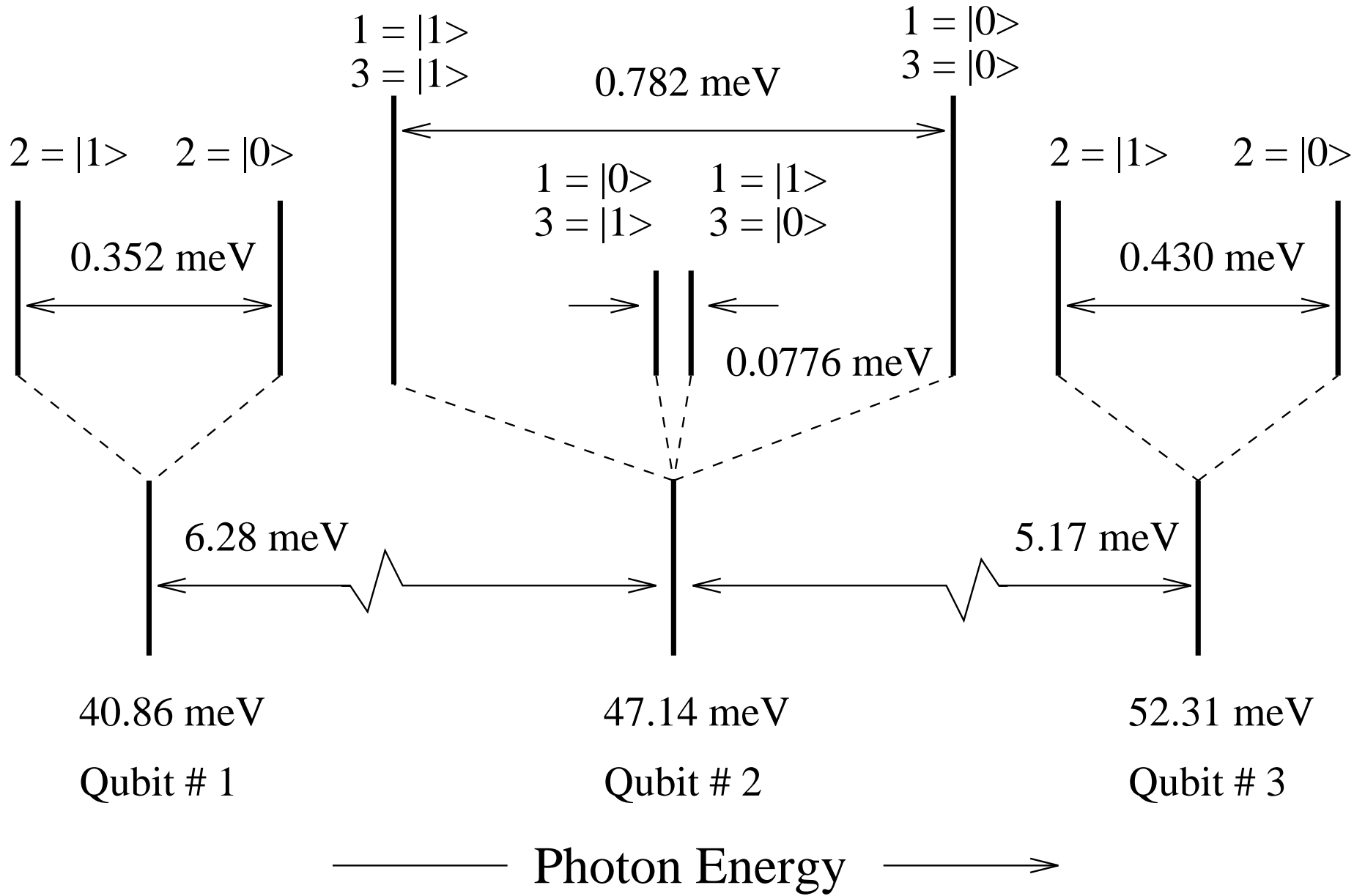
\end{document}